\documentclass{PoS}
\usepackage{epsfig}

\title{Theoretical strategies for epsilon$\,^\prime$/\,epsilon}

\ShortTitle{Theoretical strategies for epsilon$\,^\prime$/\,epsilon}

\author{\speaker{Norman Christ} %\thanks{A footnote may follow.}\\
        Columbia University, USA\\
        E-mail: \email{nhc@phys.columbia.edu}}
\author{RBC and UKQCD Collaborations}

\abstract{We review the current status of calculations of the 
two pion decays of the kaon using the first-principles methods 
of lattice gauge theory and the significant challenges that 
these calculations pose.  While a calculation with controlled 
errors at even the 10-20\% level has not yet been performed, 
present results suggest that such a calculation of the real and 
imaginary parts of the $\Delta I = 3/2$ amplitude should be 
accomplished within the next two years.  The more difficult 
$\Delta I = 1/2$ amplitude may also be now within reach.}

\FullConference{2009 KAON International Conference KAON09,\\
		 June 09 - 12 2009\\
		 Tsukuba, Japan}

\begin{document}
\bibliographystyle{JHEP}

\section{Introduction}

The two pion decays of the K meson have been an important topic
in particle physics for more than fifty years.  The same processes
that lead to the discovery of P, C and CP violation today hold
the promise of revealing the first insights into physics beyond
the standard model.  For such promise to be achieved, theoretical 
calculations must be realized whose precision matches those of 
the impressive experimental results presented at this conference.
The present theoretical challenge is the calculation of the 
hadronic matrix elements of four-quark operators which describe
the relevant electro-weak processes at the relatively long 
distances which characterize the initial and final kaon and
pion states.  This current focus on the effects of low energy QCD 
is the result of critical work over the past thirty years using 
perturbative methods to analyze these processes at short distance.
The result of this work is a low energy effective Hamiltonian of 
the form:
\begin{equation}
\mathcal{H}_{\rm eff}^{\Delta S = 1} = \frac{G_F}{\sqrt{2}}  
		\left\{ \sum_{i=1}^{10} \left[V_{ud} V^*_{us}\; z_i(\mu)
                   - V_{td} V_{ts}^*\; y_i(\mu) \right] Q_i \right\}
\label{eq:H_eff}
\end{equation}
where we use the notation of Ref.~\cite{Blum:2001xb}.  In 
Eq.~\ref{eq:H_eff} the top, bottom and charm quarks are treated as 
heavy and their effects have been incorporated using QCD perturbation 
theory so that the four-quark operators $Q_i$ contain only three 
light quarks $u$, $d$ and $s$.

Thus, the problem which this talk addresses is the calculation
of the matrix elements of $\mathcal{H}_{\rm eff}^{\Delta S = 1}$
between an initial kaon state and a final state of two pions
in either the I=0 or I=2 state: 
$\langle \pi \pi(I)|\mathcal{H}_{\rm eff}^{\Delta S = 1}|K\rangle_{I=0,2}$.
In principle, lattice QCD is ideally suited for such a calculation.
Euclidean space lattice techniques should permit the calculation of 
such matrix elements without {\it ad hoc} assumptions and with 
numerical control of all errors.  We will now discuss the 
difficulties posed by such calculations and the techniques which 
are expected to overcome them.

\section{Operator mixing and renormalization}

The 4-quark operators $Q_i$ which appear in Eq.~\ref{eq:H_eff} are
each of dimension six.  They are linear combinations of seven
independent operators, which naturally divide into three groups:
a single $(27,1)$ operator, two $(8,8)$ operators and four $(8,1)$
operators.  The operators within each group mix under 
renormalization as well as with operators of lower dimension.
Such a renormalization pattern is well understood for continuum
operators and can be managed using standard techniques if a chirally
invariant regulator, such as dimensional regularization is employed.

For a lattice calculation, these operators must now be defined using 
a lattice regulator.  Given the essential role played by chiral
symmetry in limiting the number of operators which can appear and
their mixing, it is essential to use a lattice fermion formulation
which respects chiral symmetry.  The domain wall fermion (DWF) formulation
has the needed chiral symmetry with violations that can be made
arbitrarily small as size in the fifth dimension ($L_s$) is increased.
This formulation was used in the first complete quenched calculations 
of the $K \rightarrow |0\rangle$ and $K \rightarrow \pi$ matrix
elements of these ten operators $Q_i$.~\cite{Blum:2001xb, Noaki:2001un}.
Now the RBC/UKQCD and LHPC collaborations have created substantial
ensembles for 2+1 flavor QCD for a variety of light quark masses 
and two lattice spacings.  These configurations combined with new 
larger volume, larger lattice spacing configurations will provide an 
excellent foundation for a correctly unitary, full QCD calculation 
of these decay amplitudes.

Even with such a chiral lattice formulation, we must still relate 
the lattice operators with the equivalent continuum operators 
appearing in Eq.~\ref{eq:H_eff}.   Such a matching between 
continuum and lattice operators can be accurately carried out 
using the regularization-independent, Rome/Southampton RI/MOM 
scheme, which can be applied to both continuum and lattice operators.  

The strength of the RI/MOM scheme is that it can be applied to 
lattice operators non-perturbatively by imposing simple conditions 
on off-shell gauge-fixed Greens functions evaluated using standard 
lattice gauge theory techniques.  In fact, these techniques have 
been successfully used to transform matrix elements of the bare 
lattice operators $Q_i^{\rm lat}$ into those of RI/MOM-normalized 
operators in both quenched \cite{Dawson:1999yx,Blum:2001xb} and 
full QCD \cite{Li:2008kc} calculations.  This method is increasingly 
well understood \cite{Aoki:2007xm}, with improved techniques giving 
statistical errors on the percent level.  At present the largest 
errors, on the 5\% level, come from the use of perturbation theory 
to connect the RI/MOM and $\overline{\mbox{MS}}$ schemes.  We 
conclude that the problems of operator normalization and mixing 
are under adequate control and pose no special difficulties for 
the topic at hand.

\section{Quadratic divergence}

A cause for concern when considering a calculation of the 
$\Delta I = 1/2$ $K \rightarrow \pi\pi$ amplitude is the presence 
of quadratically divergent terms.  In a lattice calculation such 
terms are finite but larger than the physical amplitude by a factor 
proportional to $1/a^2$.  While such terms will not contribute to 
any properly constructed physical amplitude, their removal, either 
by explicit subtraction  or by the averaging of the relevant matrix 
element to zero may increase statistical or to amplify systematic 
errors.  

\begin{figure}[ht]
    \hfill
    \begin{minipage}{0.45\textwidth}
%        \hspace{-0.8cm}
        \epsfig{file=O6PQS_vacsub_subtd_mres.eps,
                       width=0.9\linewidth,angle=0}
    \caption{Matrix elements of $Q_6$ (circles) which contain a 
    quadratic divergence, the subtraction term (squares) 
    and their difference (diamonds).}
    \label{fig:subtraction}
    \end{minipage}
    \hfill
    \begin{minipage}{0.45\textwidth}
%        \hspace{-1.0cm}
        \epsfig{file=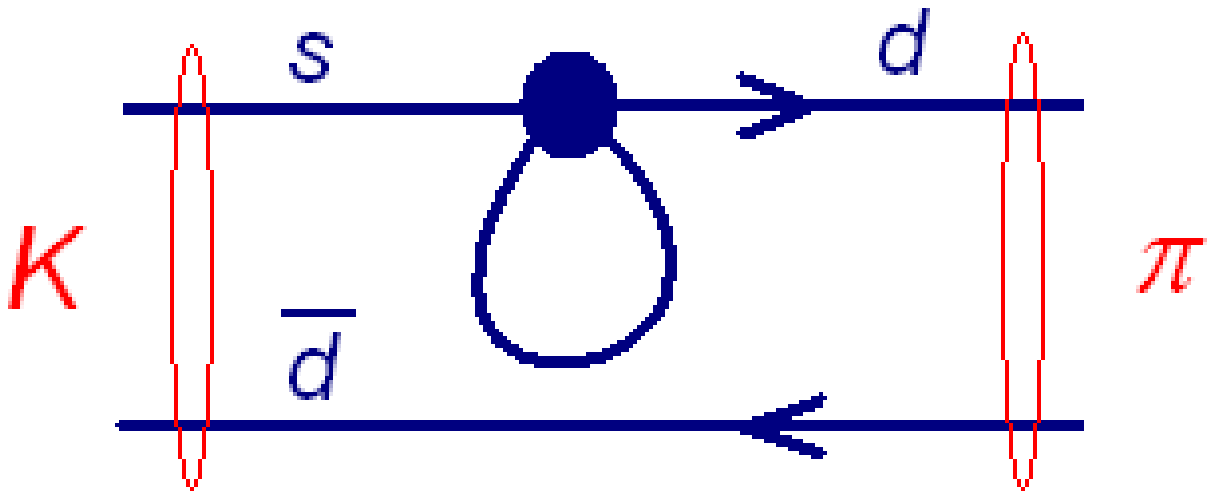,width=1.0\linewidth,angle=0} 
        \vskip 0.66in
    \caption{Example graph containing a quadratically divergent 
    quark loop which appears in the $\langle \pi|Q_6|K\rangle$
    matrix element.}
    \label{fig:penguin}
    \end{minipage}
\hfill
\vspace{-0.4cm}
\end{figure}

For example, in the standard calculation of the matrix element
of the operator $Q_6$ using leading order chiral perturbation 
theory (LO ChPT) a combination of $K \rightarrow \pi$ and 
$K \rightarrow |0\rangle$ amplitudes is required in which the 
later can be viewed as subtracting the quadratic divergence from 
the former.  These two amplitudes and their much smaller difference
is shown in Fig.~\ref{fig:subtraction}.  As can be seen this 
subtraction reduces the amplitude by nearly a factor of ten but 
still gives a result with few percent statistical errors.  These 
large matrix elements also enhance the systematic errors 
associated with finite $L_s$.  However, it can be shown 
\cite{Christ:2007zz,Sharpe:2007yd} that such errors are at most 
on the few percent level.

These quadratic divergences resulting from quark loops of the sort 
shown in Fig.~\ref{fig:penguin} appear multiplying operators 
$\overline{s}(1\pm\gamma^5) d$ which vanish when evaluated in 
momentum conserving $K \rightarrow \pi\pi$ matrix elements.  
However, such terms vanish only after an average over gauge 
configurations and therefore may introduce large statistical 
fluctuations.  If necessary, an explicit subtraction can be 
introduced which will not affect the average but should reduce the 
variance to the acceptable level found in the $K \rightarrow \pi$ 
matrix elements in Fig.~\ref{fig:subtraction}.  Thus, experience 
with $K \rightarrow \pi$ calculations \cite{Blum:2001xb,Noaki:2001un} suggests that these $O(1/a^2)$ components of the $\Delta I = 1/2$ 
operators will not pose serious difficulties.

\section{Two pion final state}

An important problem associated with such $K \rightarrow \pi\pi$
calculations is summarized by the Maiani-Testa theorem which points 
out that the large Euclidean time limit used in lattice QCD to 
project onto physical states will yield a $\pi-\pi$ state with zero 
relative momentum, not the physical state which should appear in 
the matrix element of interest.  There are now two methods to 
circumvent this difficulty.  The first uses $SU(3)\times SU(3)$ 
ChPT to relate the $K \rightarrow \pi\pi$ amplitudes of interest 
to simpler $K \rightarrow \pi$ and $K \rightarrow |0\rangle$ matrix 
elements.  This avoids dealing directly with a state containing 
two pions but, as is discussed in Section~\ref{sec:ChPT} below, 
relies on ChPT in a region where its validity is highly uncertain.  
The second more promising method is based on finite volume 
techniques~\cite{Lellouch:2000pv}.  This second approach is 
discussed below in Section~\ref{sec:finite_vol}.

\subsection{Chiral perturbation theory}
\label{sec:ChPT}

The first quenched calculations of the complete 
$K \rightarrow \pi\pi$ amplitude were carried out using LO ChPT 
\cite{Blum:2001xb, Noaki:2001un}.  While the results for the
real parts of $A_0$ and $A_2$ may have been encouraging, the
value of $\epsilon^\prime/\epsilon$ was near zero and slightly
negative---far from the experimental value.  However,
Golterman and Pallante~\cite{Golterman:2001qj} discovered that 
the ChPT structure of the quenched and full theory was very 
different, with the quenched theory possessing more singular 
chiral logarithms than are present in the complete, unquenched 
theory.

Such problems with the quenched approximation can be avoided 
by moving to full QCD simulations and the RBC/UKQCD collaboration 
has repeated the earlier quenched work using 2+1 flavor, DWF 
gauge ensembles~\cite{Li:2008kc}.  In addition to the inclusion 
of fermion loops, this new calculation explores lighter
quark masses, including ``partially quenched'' amplitudes with 
unequal valence and sea quark masses, allowing a detailed 
comparison with ChPT.  Unfortunately, the RBC/UKQCD results for 
the quark mass behavior of the standard meson masses and decay 
constants~\cite{Allton:2008pn, Boyle:2009zz} and these detailed 
studies of the $K\rightarrow \pi$ and $K \rightarrow |0\rangle$ 
amplitudes~\cite{Li:2008kc} strongly suggest that ChPT does not 
provide a reliable description for masses as large as that of 
the physical K meson.

\begin{figure}[ht]
    \hfill
    \begin{minipage}{0.45\textwidth}
        \centering
        \epsfig{file=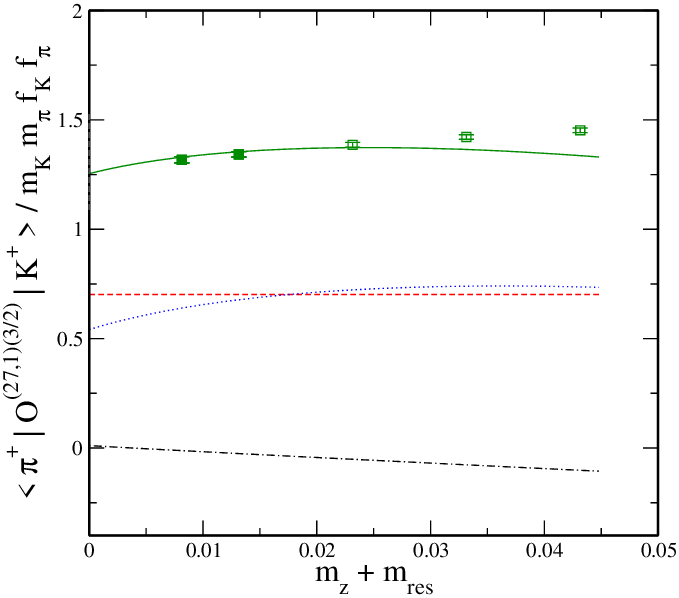,
                       width=0.9\linewidth,angle=0}
    \caption{A NLO ChPT fit to the ratio in 
    Eq.~\protect\ref{eq:O27_NLO} compared to the lightest
    sea quark results.  The individual components of the fit 
    as well as their total are shown.  The NLO analytic and
    logarithmic terms appear in the middle while the smallest
    contribution, near zero, is the LO term.}
    \label{fig:O27}
    \end{minipage}
    \hfill
    \begin{minipage}{0.45\textwidth}
        \centering
        \epsfig{file=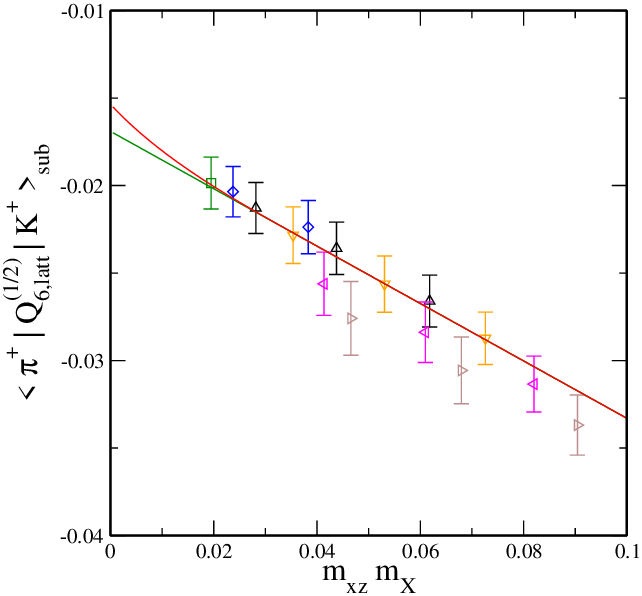,width=0.9\linewidth,angle=0}
    \caption{Results for the $(8,1)$ operator $Q_6$ after subtraction
    of the quadratic divergence together with a LO chiral fit.  The
    small deviation shown at small quark masses is an estimate of
    the possible effect of a NLO chiral logarithm.}
    \label{fig:O6_LO}
    \end{minipage}
    \hfill
%\vspace{-0.4cm}
\end{figure}

In Fig.~\ref{fig:O27} we show an attempt to fit next leading
order (NLO) ChPT to the ratio 
\begin{equation}
\langle\pi|O^{(27,1)}|K\rangle/f_K f_\pi m_\pi m_K
\label{eq:O27_NLO}
\end{equation}
in order to extract the LEC $\alpha_{27}$.  This figure reveals two difficulties.  First, we are fitting the ratio shown in 
Eq.~\ref{eq:O27_NLO} because we were not able to obtain a sensible 
fit to the simpler matrix element in the numerator.  One might 
argue that the NLO ChPT terms in this ratio will be reduced 
because of significant cancelation between the NLO chiral 
logarithms which appear in the numerator and denominator.  
However, a circumstance in which we must construct artificial 
ratios in order to obtain sensible fits naturally raises serious 
doubts about the applicability of ChPT.

The second problem is the dominance of the next leading order (NLO)
term over that of leading order.  While the leading order term 
may be accidentally suppressed, the failure of the fit to show 
a standard hierarchy among the orders of the expansion suggests 
the more probable situation that ChPT is inapplicable and the 
result is a fit to an essentially arbitrary function with no 
ordering between the various terms.  A similar situation is 
found for the other two $\Delta I = 3/2$, $(8,8)$ operators.

The situation is less clear for the (8,1) $\Delta I = 1/2$ operators
because the NLO ChPT contains more low energy constants (LEC's) 
making our data inadequate to carry out a complete NLO ChPT fit.  As
can be seen in Fig.~\ref{fig:O6_LO} a simple linear fit (leading 
order in ChPT) describes the data very well.  However, as shown by 
the additional up-turning curve added to the left of the data 
points, possible NLO logarithmic behavior that is entirely 
consistent with our calculated points can change the slope at 
vanishing light quark mass by a factor of two.  Since it is this 
slope which is the LEC of interest, $\alpha_6$, we must assign a 
100\% systematic error to our result.  

In summary, while the use of the 2+1 flavor DWF configurations 
has removed the uncertainties associated with the quenched 
approximation, the use of smaller quark masses and partial 
quenching has raised new concerns about the validity of ChPT in 
the kinematic region needed to determine the $K \rightarrow \pi\pi$
amplitudes.  First, as discussed above, our calculation 
suggests that the needed LEC's cannot be reliably determined
from our present range of masses.  This problem can be addressed 
by moving to lighter masses.  However, there is a second, more
serious difficulty.  Even if these LECs were known, we would 
still require a second application of ChPT to compute the 
$K \rightarrow \pi\pi$ amplitudes.  This use of ChPT would involve
exactly the region of large quark masses and pion momenta
where the studies above suggest the theory fails.  There is
no solution to this problem if we chose to work in the physical
world with 498 MeV kaons.

\subsection{Calculation with two pion final states}
\label{sec:finite_vol}

We conclude that it is necessary to directly compute matrix
elements using states containing two pions.  Evaluating matrix
elements of such a 2-particle state using Euclidean space 
lattice techniques is now well understood~\cite{Lellouch:2000pv}
but still presents serious practical challenges.  The key to
computing such $\pi-\pi$ matrix elements is to understand and 
exploit the finite volumes that necessarily appear in a lattice
calculation.  

While the underlying weak operator connects the initial K meson 
to two pions in an s-wave, the rectangular box of a lattice 
calculation mixes that $l=0$ state with states with $l = 4, 8, ....$   
In the reasonable approximation that the $\pi-\pi$ phase shifts,
$\delta_l$, vanish for $l>0$, then the weight of the $l=0$ component 
of the finite volume eigenstates can be computed knowing only 
$\delta_{l=0}$ and allowing the physical $l=0$ matrix element to 
be extracted from that of the finite volume state.  While the 
finite volume $\pi-\pi$ ground state is close to threshold with 
nearly zero relative momentum, there are a series of excited 
states with relative pion momenta which differ from the free-particle
multiples of $2\pi/L$ by calculable amounts, again determined by 
$\delta_0$.  

For a cubic box with $L=6$ fm, the pions in the first excited 
state have relative momentum very close to the physical value of 
$p=205$ MeV.  With present resources, this is an inaccessibly
large volume.  However, this problem can be circumvented by
three strategies.  The first starts with a K meson with non-zero
momentum.  Momentum conservation requires that the two pion state
carry this same momentum which typically implies that one pion 
must remain at rest.  If the initial kaon has a momentum of
740 MeV, then the relative momentum of the two final pions is
physical.  This 740 MeV value can be easily achieved on a practical
3 fm lattice if $p_K = (1,1,1)2\pi/L$ or $p = \sqrt{3}\,2\pi/L$.
Of course, with all methods, the challenge of an $L=6$ fm volume
must eventually be met as the mass of the pions used approaches 
its physical value.   The rule-of-thumb that $L m_\pi \ge 4$
to avoid finite volume effects also requires $L \ge 5.7$fm
when $m_\pi = 138$ MeV. 

This $p_K >0$ approach has been studied by T. Yamazaki
\cite{Yamazaki:2008hg} for the $\Delta I = 3/2$ amplitudes with 
encouraging results.  However, the large momentum of the $K$ and 
$\pi$ is a cause for concern.  Since momentum conservation takes 
effect only after the average over gauge fields has been performed, 
the amplitude on each configuration will be dominated by the
much larger contributions from $K$ and $\pi$ states at rest.  The 
large cancelation needed to give the correct averages implies a 
corresponding large statistical noise.  Never-the-less, this method 
works for both the $\Delta I = 3/2$ and $\Delta I = 1/2$ cases 
and for $\Delta I = 1/2$ automatically removes the vacuum
contribution which necessarily carries zero momentum.

A second approach imposes anti-periodic boundary conditions on 
the pions.  This is easily done for the $\Delta I = 3/2$ amplitude 
with its $I=2$ final state \cite{Kim:2005gk} by using anti-periodic 
boundary conditions for the $u$ quark but periodic boundary conditions 
for the $d$.  Isospin symmetry implies that the $A_2$ amplitude can be 
determined from a matrix element with a $|\pi^+\pi^+\rangle$ final state, where both $\pi^+$ mesons obey anti-periodic boundary conditions and 
therefore carry the physical momentum $\pi/L = 205$ MeV when $L = 3$ fm, 
a practical requirement.  For $I=2$, the final state must be
constructed from valence quarks and one can argue
\cite{Sachrajda:2004mi} that if the anti-periodic boundary conditions 
are applied only to the valence quarks in the calculation, 
allowing standard periodic boundary condition gauge configurations,
then only errors exponentially suppressed by the lattice size 
are introduced.

Imposing anti-periodic boundary conditions on the pions is more 
difficult for the $I=0$ state where the quark content is less 
controlled.  An attractive method is to impose G parity boundary 
conditions \cite{Kim:2002np} on the quarks which guarantees 
anti-periodic pions:
\begin{equation}
\left( 
  \begin{array}{c} u(x) \\ d(x)  \end{array}
\right) \rightarrow \left( 
  \begin{array}{c} {\cal C} \overline{d}(x+L \hat e_i) \\
                  -{\cal C} \overline{u}(x+L \hat e_i) \end{array}
\right) \rightarrow \left( 
  \begin{array}{c} -u(x+2L \hat e_i) \\
                   -d(x+2L \hat e_i) \end{array}
\right)
\end{equation}
where ${\cal C}$ is the standard $4 \times 4$ charge-conjugation 
matrix and $\hat e_i$ a unit vector in a direction in which these
G parity boundary conditions are imposed.

These boundary conditions require special gauge ensembles where 
the gauge fields obey charge conjugate boundary conditions and 
the light sea quarks also obey G parity boundary conditions.  
In addition, special treatment is required for the strange quark 
which is best made part of a fictitious iso-doublet.  These sea 
quarks can be represented by standard, positive-definite DWF 
determinants (no Pfaffians needed) although a square root of the 
strange quark determinant is required to keep the correct number 
of flavors.  Given the computational costs required by the 
disconnected diagrams, the extra difficulty of generating these 
special gauge configurations may be relatively minor, making G 
parity boundary conditions an attractive approach to the 
$\Delta I = 1/2$ amplitude.

\section{Disconnected diagrams}

Because the $I=0$, $\pi -\pi$ state has the flavor quantum
numbers of the vacuum, $I=0$, $\pi -\pi$ propagators and
$K \rightarrow \pi\pi(I=0)$ amplitudes all contain diagrams
in which no quark lines connect the initial and final states.
These quantities then contain a vacuum contribution which does 
not vanish as the separation between the source and the sink 
or weak decay operator grows.  Such unphysical vacuum terms 
either vanish because of inconsistent non-flavor quantum numbers 
({\it e.g.} $\vec p \ne 0$) or must be explicitly subtracted.  
Unfortunately, even when removed, the separation-independent 
fluctuations of such vacuum terms can quickly overwhelm 
the exponentially decreasing physical signal, presenting major 
problems for the lattice calculation of many interesting 
physical quantities.

\begin{figure}[ht]
    \hfill
    \begin{minipage}{0.45\textwidth}
        \hspace{-1.0cm}
        \epsfig{file=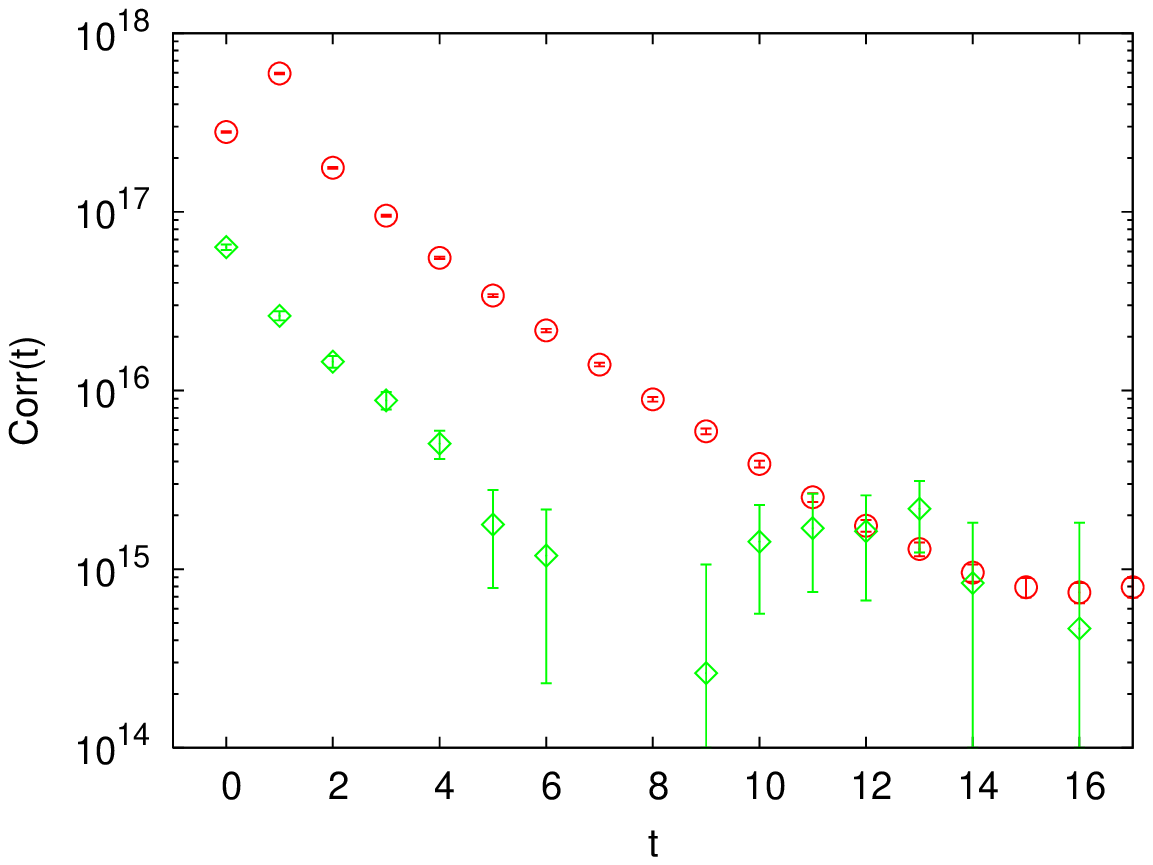,width=1.1\linewidth,angle=0} 
    \end{minipage}
    \hfill
    \begin{minipage}{0.45\textwidth}
        \hspace{-0.8cm}
        \epsfig{file=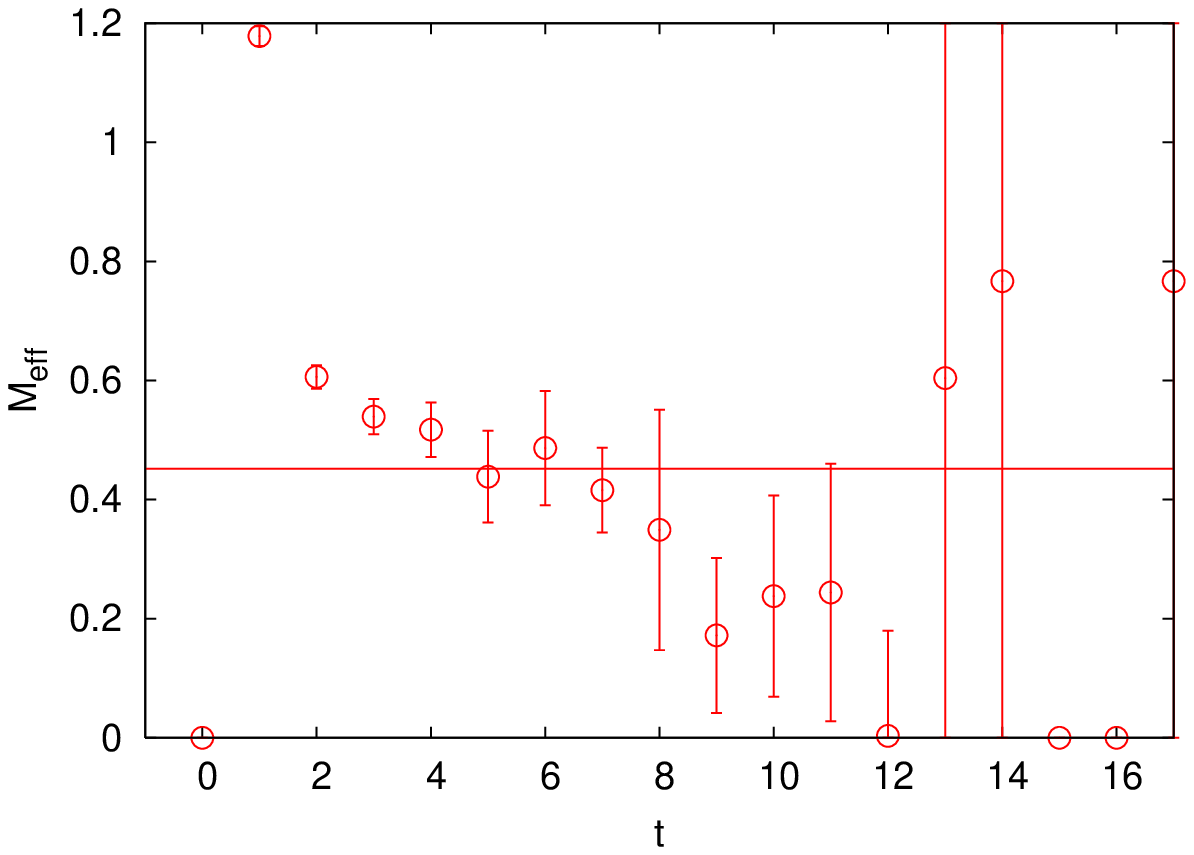,width=1.1\linewidth,angle=0}
    \end{minipage}
\caption{The left panel shows the connected (circles) and disconnected 
(diamonds) contributions to the $I=0$ $\pi-\pi$ correlator.  The 
right-hand panel shows the resulting 2 pion effective mass.  The 
missing points in the left panel result from negative values for 
the amplitude which cannot be shown on a logarithmic plot.}
\label{fig:disconnected}
\hfill
%\vspace{-0.4cm}
\end{figure}

These difficulties may be least severe for the $I=0$, $\pi -\pi$ 
system.  As the quark masses become more physical and the pions 
less massive, the $\pi -\pi$ signal falls less rapidly
with increasing time allowing a signal to be obtained at 
larger times.  In Fig~\ref{fig:disconnected} we show preliminary 
results of $I=0$, $\pi -\pi$ scattering study being carried out
by Qi Liu.  For small time separations the disconnected amplitude 
can be determined and appears much smaller than the connected 
piece.  However for times of 5 or greater, the noise in the 
disconnected part begins to contribute substantially to the 
error in the complete amplitude, severely limiting the temporal 
region where a plateau in the effective mass and the needed 
evidence for asymptotic behavior can be established.

While this preliminary work is only beginning to uncover the
difficulties of dealing with the disconnected contributions to the 
$K \rightarrow \pi\pi(I=0)$ amplitudes, we believe there is  
reason for optimism.  These results were obtained in a few months
on 1024-node QCDOC partitions.  One hundred and thirty 
$16^3\times 32$ lattice configurations were analyzed computing
32 sets of propagators on each, using wall sources located on each 
possible time hyperplane.  With larger computer resources, working 
with larger lattice volumes and collecting larger statistics will 
be practical.  While this will not permit the analysis to be 
extended to significantly larger times, it will yield more accurate 
results in the time range $0 \le t \le 6$ where we hope that 
the introduction of more interpolating operators will allow the 
extraction of both the ground and excited states so that physical 
matrix elements can be determined without relying on a simple 
large-time limit to project onto the ground state.

\section{Outlook}

Substantial efforts over the past ten years strongly suggest that
a lattice calculation of the complex $K \rightarrow \pi\pi$ 
amplitudes $A_0$ and $A_2$, accurate to 10-20\%, requires a full 
(unquenched) lattice QCD calculation in which on-shell two pion
states are studied.  This experience suggests that issues of
divergent penguin graphs and operator renormalization and mixing
can be treated with a few percent precision using chiral fermions 
and non-perturbative RI/MOM techniques.  Existing finite volume
methods should permit the direct evaluation of matrix elements
with on-shell $|\pi\pi\rangle$ states.  The disconnected diagrams
which appear in the $I=0$ amplitude pose the most serious
challenge which may be overcome with substantial statistics and
a careful multi-state analysis of correlation functions evaluated
at relatively small time separations.

Based on this assessment, computing the complex $K \rightarrow \pi\pi$ 
amplitudes has become an important research goal of the RBC and
UKQCD collaborations.  The first objective is the calculation
of the $\Delta I = 3/2$ amplitude $A_2$ which can be done on standard,
2+1 flavor gauge configurations using anti-periodic valence quarks.  
We are presently carrying out a quenched calculation on a 3.6 fm, 
$24^3 \times 64$ volume~\cite{Lightman:2009ka}.  We expect this to 
provide interesting results for matrix elements of the three 
$\Delta I = 3/2$ operators and guidance for a follow-on calculation 
using the 2+1 flavor, 4.6 fm lattice ensembles now being generated by 
the RBC/UKQCD and LHPC collaborations on the ALCF at Argonne.  We 
expect this effort to determine $A_2$ to an accuracy of 10-20\% 
within two years.

For the $\Delta I = 1/2$ amplitudes a meaningful quenched calculation 
is not possible and our initial calculations are being performed
on 2+1 flavor $16^3 \times 32$ lattice volumes with 430 MeV pions.  
A first, complete $K \rightarrow \pi\pi$ calculation is now 
underway including all connected and disconnected diagrams.  With 
the experience gained in this calculation, we hope to exploit the 
next generation of 100-teraflops sustained computers as they become 
available in roughly one year's time to move to larger volumes and 
lighter pion masses.  This may allow calculations of both the 
$\Delta I = 3/2$ and 1/2 amplitudes giving 10-20\% results for 
both the $\Delta I = 1/2$ rule and $\epsilon^\prime/\epsilon$ in 
2-3 years.  While this may easily be optimistic, it is surely an 
exciting and increasingly realistic goal to pursue.

\section*{Acknowledgments}

The author acknowledges the many ideas and insights of 
his RBC/UKQCD collaborators which have been summarized here and 
the work of Shu Li, Matthew Lightman and Qi Liu whose results were 
described above.  This research was supported by DOE grant 
DE-FG02-92ER40699.

%\bibliography{kaon09}

\providecommand{\href}[2]{#2}\begingroup\raggedright\endgroup

\end{document}